\documentclass[twocolumn,showpacs,preprintnumbers,floatfix,pre]{revtex4}
\usepackage{graphicx}
\usepackage{amsmath} 
\usepackage{color}

\begin{document}

\title{Compaction dynamics of a magnetized powder}

\author{G. Lumay$^{\dag,\ddag}$, S. Dorbolo$^{\dag,\ddag}$ and N. Vandewalle$^{\dag}$}
\affiliation{\dag GRASP, Universit\'e de Li\`ege, B-4000 Li\`ege, Belgium.\\
\ddag F.R.S.-FRNS, B-1000 Bruxelles, Belgium}

\begin{abstract}
We have investigated experimentally the influence of a magnetic interaction between the grains on the compaction dynamics of a granular pile submitted to a series of taps. The granular material used to perform this study is a mixture of metallic and glass grains.  The packing is immersed in an homogeneous external magnetic field. The magnetic field induces an interaction between the metallic grains that constitutes the tunable cohesion. The compaction characteristic time and the asymptotic packing fraction have been measured as a function of the Bond number which is the ratio between the cohesive magnetic force and the grain weight. These measurements have been performed for different fractions of metallic beads in the pile. When the pile is only made of metallic grains, the characteristic compaction time increases as the square root of the Bond number. While the asymptotic packing fraction decreases as the inverse of the Bond number. For mixtures, when the fraction of magnetized grains in the pile is increased, the characteristic time increases while the asymptotic packing fraction decreases. A simple mesoscopic model based on the formation of granular chains along the magnetic field direction is proposed to explain the observed macroscopic properties of the packings.\pacs{81.05.Rm,81.20.Ev} 
\end{abstract}

\maketitle


\section{Introduction}

Over the last decade, the compaction of granular material has been the subject of numerous studies in the physics community \cite{deGenne,Jaeger,Kudrolli2004} as the problem concerns a large panel of both industrial and fundamental research.  The majority of these studies are focused on cohesiveless granular materials. However, a better fundamental knowledge of cohesive powder properties is required for their manipulation, in particular within the recent development of nanopowder technology. Indeed, cohesive forces are known to strongly affect the flow and the static  properties of fine powders, since they induce the formation of large aggregates \cite{Lumay2008}.

The compaction dynamics of cohesiveless granular material submitted to a series of taps has been studied experimentally, theoretically and numerically \cite{densFluc,densityRelax,richard,kww1,LumayVandewalle2,Tetris,Tetris2,slowRelax,boutreux}. Different laws have been proposed for the evolution of the volume fraction $\eta$ as a function of the number $n$ of taps. Among others, one has proposed the inverse logarithmic law \cite{densFluc} 
\begin{equation}
\label{eq:logLaw}
\eta(n) = \eta_{\infty} - \frac{\Delta \eta}{1 + B \; \ln(1+\frac{n}{\tau})}
\end{equation} where the parameters $\eta_{\infty}$ and $\Delta \eta$ are respectively the asymptotic volume fraction and the maximum variation of the volume fraction. The dimensionless parameter B depends on the acceleration during each tap and $\tau$ is the characteristic relaxation time for the grain reorganization process.

The difference between a powder and a granular assembly is the increase of the importance of cohesion forces in respect to the gravitation because the grains are much smaller in powder than in granular assembly.  It is difficult to imagine a model powder for which it is possible to tune the size or the shape of the grains independently to the cohesion.  In order to study the influence of the cohesive forces on the properties of granular assemblies, several methods can be used. For example, one may change the size of the particles \cite{Castellanos}. However, controlling the size and the shape of particles is a difficult task in particular for small cohesive objects. Liquid bridges can also be used to increase the interparticle forces \cite{Kudrolli2008} and, under certain conditions, to lubricate contacts \cite{Kudrolli2007} in addition to cohesive forces. 
\begin{figure}[h]
\begin{center}
\includegraphics[scale=0.4]{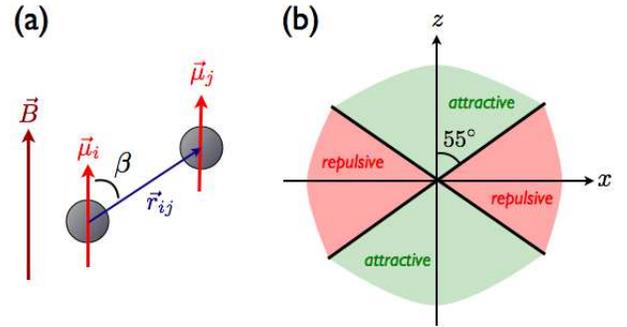}
\end{center}
\caption{(a) Interactions between two ferromagnetic particles in a magnetic field $\vec B$. The $\vec{\mu}$ vector denotes the magnetic dipole induced by the magnetic field $\vec B$, $\vec{r_{ij}}$ denotes the vector between the dipoles. (b) The magnetization of two ferromagnetic particles in a vertical magnetic field induces repulsive interaction when $\vec{r_{ij}}$ is perpendicular to $\vec B$ and attractive interaction when $\vec{r_{ij}}$ is parallel to $\vec B$.}
\label{fig:interaction}
\end{figure}

The idea of our study is the use the magnetic force to measure the impact of the cohesion on the compaction properties of a granular assembly. The interparticle force can be tuned continuously by varying the strength of the external magnetic field $\vec B$. This technique has previously been used to study the influence of the cohesion on the packing fraction \cite{Lumay2007,Forsyth2001} and on the avalanche angle \cite{Forsyth2001_2} of granular assemblies.  The choice of the size of grains (beads) is important: not too small to avoid `natural' cohesion, not too large to avoid the use of high magnetic field and to remain in powder conditions.

The influence of the magnetic field can be sum up as following. Each ferromagnetic particle becomes a magnetic dipole characterized by a magnetic moment $\vec{\mu}$. We assume that the magnetic moments $\vec{\mu}$ are parallel to the applied magnetic field $\vec{B}$. This is an approximation when we consider many interacting particles. The potential energy $U_{ij}$ between two magnetic dipoles $i$ and $j$ separated by a distance $r_{ij}$ (see Figure \ref{fig:interaction}) is given by
\begin{equation}
U_{ij} = \frac{\mu_0 \mu_i \mu_j}{4 \pi} \frac{ 1 - 3 \cos^2\beta} {r_{ij}^3},
\end{equation} where $\beta$ denotes the angle between the vector $\vec{r_{ij}}$ and the magnetic field $\vec{B}$. As illustrated in Figure \ref{fig:interaction}, the potential is attractive when $\beta < 55^\circ$ and repulsive when $ 55^\circ < \beta < 90^\circ$. These interactions between the particles change significantly the internal structure of the packing and the force network. Moreover, due to the directional nature of the force acting between dipoles, anisotropic particle aggregates form along the field direction. 

In this paper, we study the influence of the magnetic interaction between the grains on the compaction dynamics. The size of the grains has been chosen such as the cohesion is weak regards their weights and small enough to be considered as a powder.  First, a granular material made of metallic beads has been used. Afterward, mixtures of glass and metallic beads have been considered. The compaction characteristic time $\tau$ and the asymptotic packing fraction $\eta_{\infty}$ have been measured for different magnetic field strength and for different fraction $\phi$ of metallic beads in the mixture. At the end of the paper, a simple mesoscopic model is proposed in order to describe the properties of a magnetic packing. 


\section{Experimental setup}

\begin{figure}
  \includegraphics[scale=0.25]{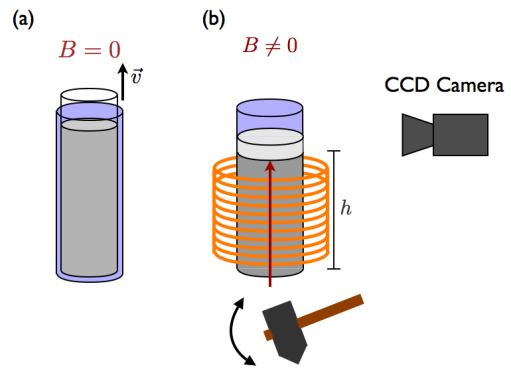}
  \caption{Sketch of the experimental setup. (a) An initialization protocol is used to obtain a reproducible and homogenous initial packing fraction $\eta_0$. A smaller bottomless tube is inserted into the main glass cylinder. This tube is filled with the granular material and is removed at constant speed $v$. (b) After the initialization process, the magnetic field $B$ is set. An electromagnetic hammer produces periodic taps on the bottom of the system. A CCD camera take a picture of the top of the pile after each tap. In order to keep the granular/air interface flat, a light cylinder is placed on the top of the pile.}
\label{fig:setup}
\end{figure}

A sketch of our experimental setup is illustrated in Figure \ref{fig:setup}. A glass tube of internal diameter $D=21$ mm is placed into a vertical magnetic coil where a constant current can be injected. The magnetic field is thus parallel to the gravity field. The strength of the magnetic field can be fixed between 0 and 150 Gauss. We checked that the variation of the magnetic field along the vertical axis of the coil does not exceed 10 \%. The average diameter of both metallic and glass grains is $d$=100 $\mu$m with a polydispersity of about 50\%. 

In order to obtain a reproducible and spatially homogeneous initial packing fraction, the following initialization protocol has been used. A narrower and bottomless glass tube (external diameter slightly below 21 mm) is inserted into the main vessel. Afterward, the small tube is filled with the granular material. The small tube is then removed upward at a low and constant velocity $v= 1$mm/s. The current in the coil is set after the initialization protocol in order to obtain a constant initial packing fraction $\eta_{0}$. A light disk is gently placed on the top of the pile in order to keep it flat during the compaction process. The system used to produce the taps under the tube is an electromagnetic hammer as described in \cite{LumayVandewalle2,Lumay2006}. During one tap, the system undergoes a short peak of acceleration (the width of the peak is 0.25 ms and the maximum intensity is 15g) and some damped oscillations during 2 ms. To obtain a compaction curve, two thousand taps are produce. Successive taps are separated by one second. Therefore, the packing relaxes after each tap.

The packing fraction of the pile is measured by image analysis from a CCD camera. The average position $h$ of the upper interface of the packing is measured after each tap and gives the packing fraction $\eta$ by using the known value for the volumic mass $\rho$ of the beads and the total mass $m$ of grains. One has $\eta = \frac{4 m}{\rho h D^2 \pi}$. The part of the pile situated above the coil is typically less than 10 \% of the total height of the pile. 

To obtain a relationship between the magnetic field strength and the inter-grain interactions, a reference magnetic field $B_{ref}$ has been measured. This  reference magnetic field gives the balance between the magnetic force acting on two contacting grains aligned with the magnetic field and the weight of one grain. Then, we can compute the Bond number $Bo$ which is the ratio between the cohesive force $F_c$ and the grain weight $mg$. If we assume that the inter-grains force is proportional to the square of the magnetic field \cite{Forsyth2001}, we obtain $Bo = B^2/B_{ref}^2$. The reference magnetic field for the metallic beads is $B_{ref} = 50$G. 


\section{The pure case: only metallic grains}
\subsection{Observations}
\begin{figure}
  \includegraphics[scale=0.55]{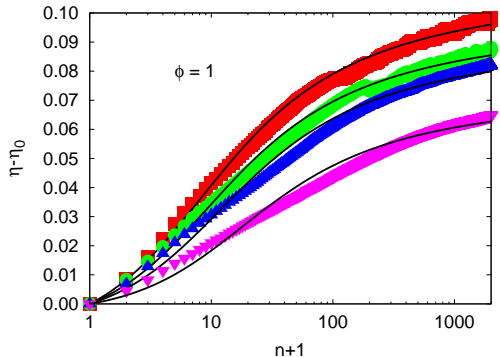}
  \caption{Typical compaction curves corresponding to different Bond numbers $Bo$. From the top curve to the bottom one, we have: $Bo = 0.0$, $Bo = 0.4$, $Bo = 1.0$ and $Bo = 5.8$. The fraction of metallic grains is $\phi=1$. The experimental data are fitted with the logarithmic law Eq(\ref{eq:logLaw}). }
\label{fig:etaVsN}
\end{figure}

\begin{figure*}
  \includegraphics[scale=0.55]{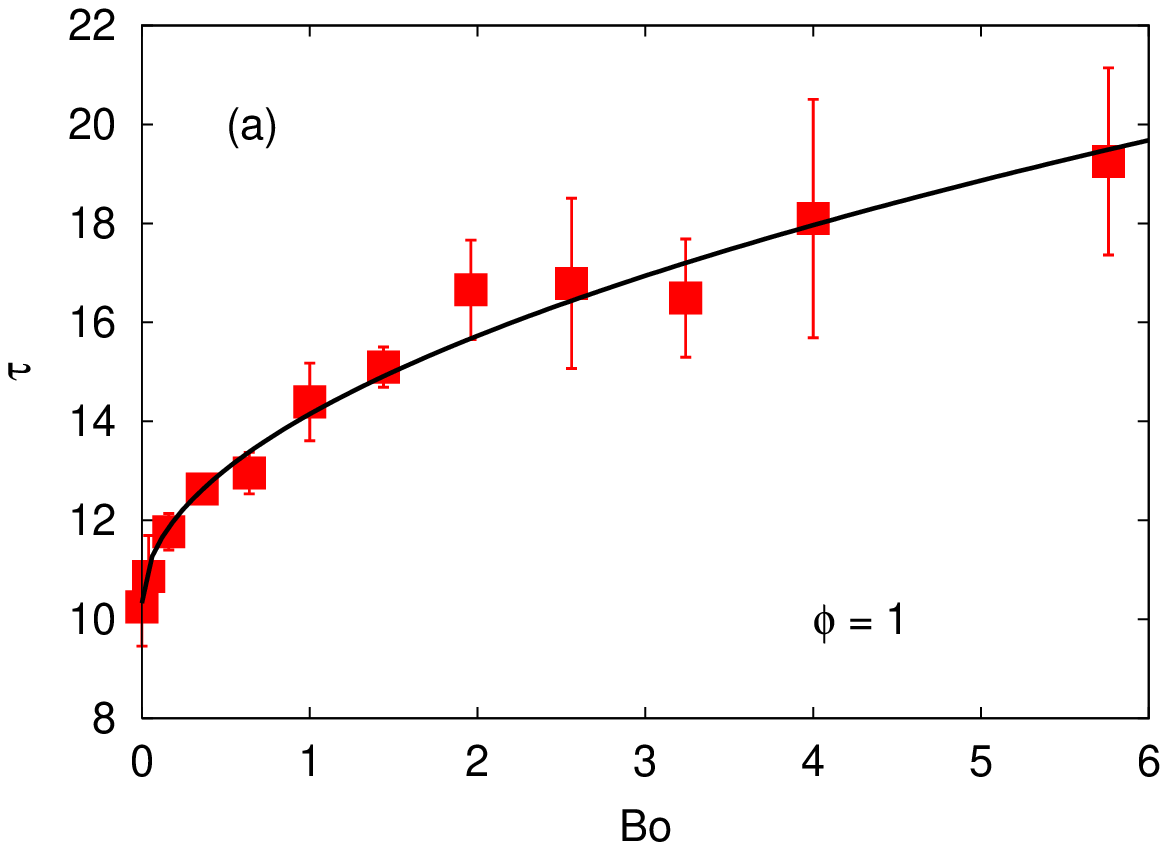} \includegraphics[scale=0.55]{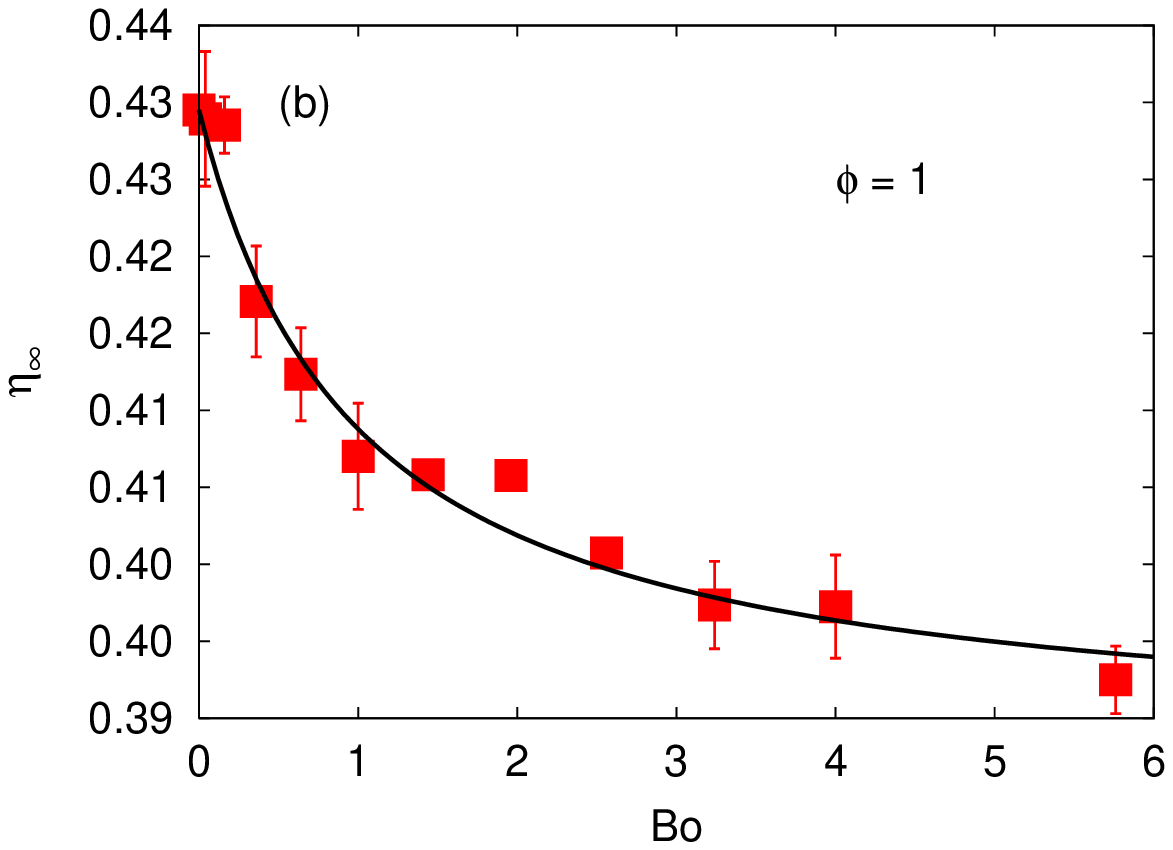}
  \caption{(a) Evolution of the characteristic time $\tau$ as a function of the Bond number $Bo$. The fraction of metallic grains is $\phi=1$. Each point corresponds to an average over three experiments. The continuous line is a fit with a square root law $\tau = \tau_0 + a \sqrt{Bo}$. The only fitting parameters is $a = 3.8 \pm 0.2$. The value of the characteristic time when $Bo = 0$ is $\tau_0 = 10.3$. (b) Decrease of the asymptotic packing fraction $\eta_{\infty}$ with the Bond number $Bo$. The fraction of metallic grains is $\phi=1$. Each point corresponds to an average over three experiments. Error bars are indicated. The continuous line is a fit by Eq. (\ref{eq:rhoInfVSBo}). The only fitting parameter is $\eta_{\infty,Bo=\infty} = 0.388 \pm 0.001$.}
\label{fig:tauVsBo}
\end{figure*}

To perform the compaction experiments, the pile is initialized with the method explained previously. The magnetic field is set after the initialization of the pile in order to obtain always the same initial packing fraction $\eta_{0}$. The figure \ref{fig:etaVsN} presents four typical compaction curves obtained with different values of the magnetic field strength. The corresponding Bond numbers are $Bo=0$, $Bo=0.4$,  $Bo=1.0$ and $Bo=5.8$. The continuous curves are the fits with the logarithmic law Eq. (\ref{eq:logLaw}). The parameter B is fixed to 1 in order to minimize the number of free fitting parameters. Indeed, the parameters $B$ and $\tau$ are known to be strongly anticorrelated.

The evolution of the characteristic time $\tau$ obtained from the fits as a function of the Bond number is presented in Figure \ref{fig:tauVsBo}. The strong increase of $\tau$ observed for a low Bond number shows that the compaction dynamics is strongly influenced by the interactions between the grains. The cohesion induced by the magnetic field slows down the compaction process. We have observed previously this effect with fine cohesive powder \cite{Lumay2006}. The evolution $\tau$ as a function of the Bond number $Bo$ is well fitted by a square root law (see Figure \ref{fig:tauVsBo}(a)). Therefore, $\tau$ increases linearly with the magnetic field strength $B$. In the cohesionless case ($Bo=0$ and $B=0$), some fluctuations are observed in the compaction curve after 80 taps (see figure \ref{fig:etaVsN}). These fluctuations are related to a convective process inside the packing. When the Bond number increases, this convective process takes place later. Moreover, for high Bond numbers ($Bo > 1$), i.e. when the cohesion between the grains is higher than the weight of one grain, this convection process disappears. 

Figure \ref{fig:tauVsBo}(b) presents the asymptotic packing fraction $\eta_{\infty}$ as a function of the Bond number $Bo$. The fraction $\eta_{\infty}$ decreases when the magnetic interaction between the grains strengthens. This behavior has been observed in a previous work \cite{Lumay2007} for the packing fraction $\eta_0$ of a metallic grains pile created in a vertical magnetic field. One should note that the asymptotic packing fraction obtained in the cohesiveless case is far lower than the random close packing limit $\eta_{RCP} \sim 0.64$. The low value of the packing fractions obtained in our experiment is due to the roughness of the metallic grains.

\subsection{Modeling}

In order to estimate the compaction characteristic time $\tau$, numerous theoretical models \cite{boutreux, Ludewig2008} have been developped. The great majority of them considers particles which are moving in ``cages" constituted of neighboring particles. As shown recently  \cite{Ludewig2008} in numerical simulations, the probability that a particle escapes a cage is related to $\tau$. The escape probability is a complex function of both packing fraction $\eta$ and some energy barrier $E_b$ related to the work for separating two grains. Roughly, one has a scaling $\tau \sim E_b$ \cite{Ludewig2008}. If one considers that the energy barrier is herein linked to the work for separating two magnetic dipoles, one obtains the scaling $\tau \sim Bo$ which is not observed in Figure \ref{fig:tauVsBo}. As a consequence, the assumption of caging dynamics does not hold in our system. 

Theoretical and numerical works have been done to predict the packing fraction $\eta$ of an assembly of particles. Thermodynamic approaches \cite{Edwards1994} and statistical methods \cite{Williams1998} have been developped. For identical spherical grains, an average over all possible Voronoi cells around particles allows to estimate the random loose packing fraction in 2d. A similar approach \cite{Vandewalle2009} allows to estimate the packing fraction of bidisperse 2d grains. Herein, the situation is more complex because the attractive and repulsive interactions should be taken into account. Note that vertical chains are favored by the magnetic field.


\begin{figure}
  \includegraphics[scale=0.4]{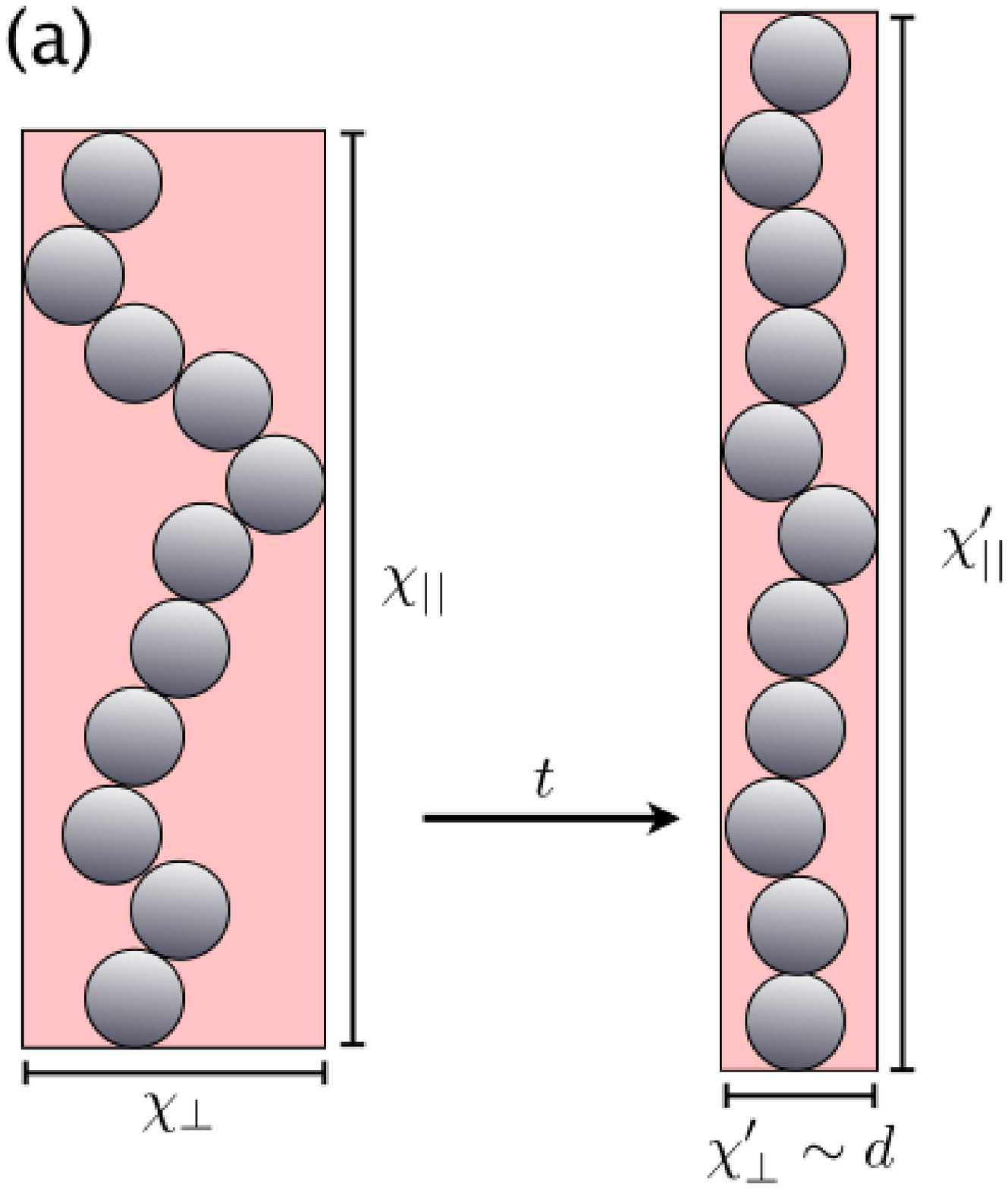}
  \includegraphics[scale=0.4]{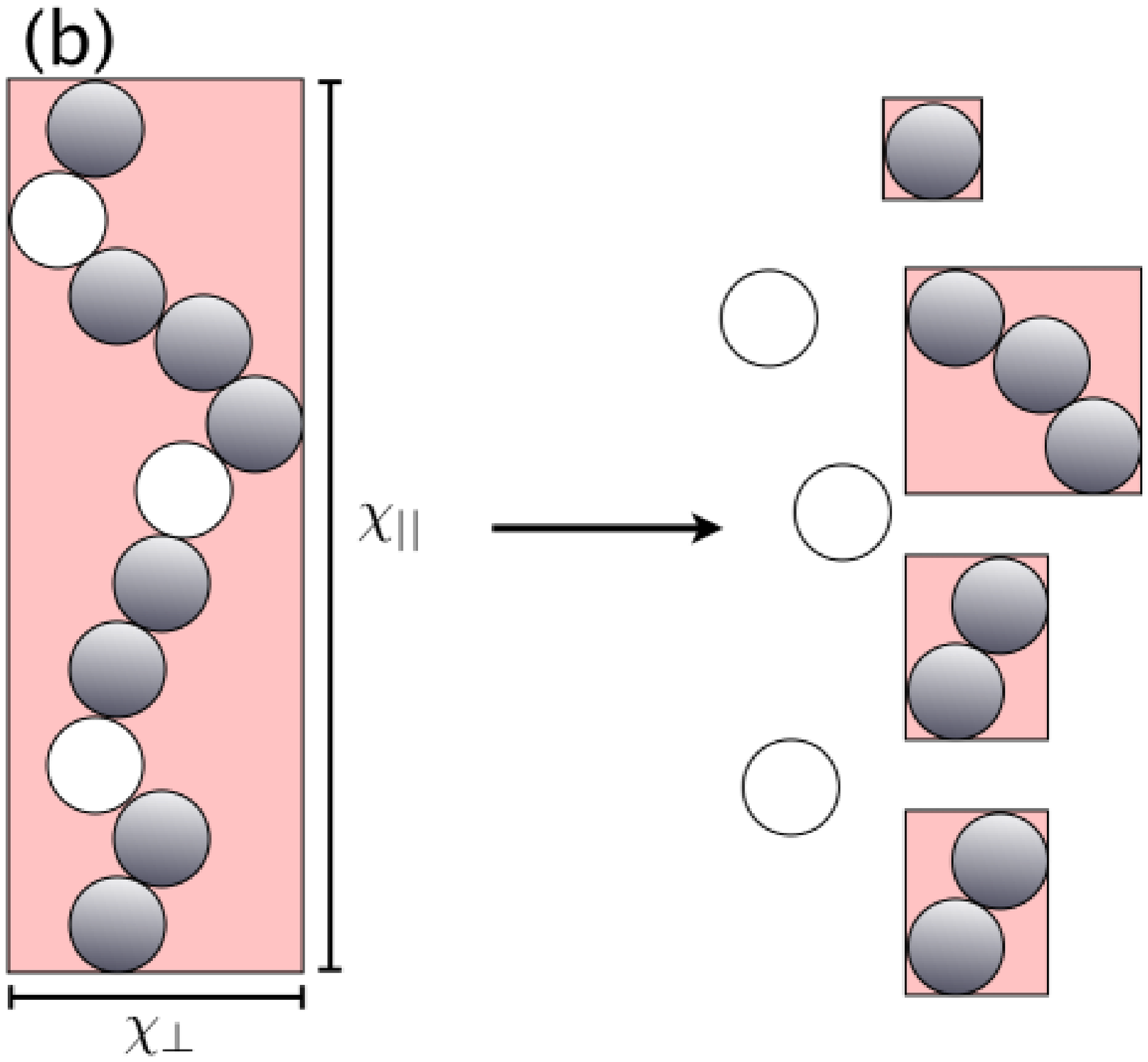}
  \caption{(top) An aggregate composed by several metallic beads before and after the compaction process. The disordered nature of the chain formation defines two typical length $\chi_{||}$ and $\chi_{\perp}$ which are related to each other (see text). (bottom) When the fraction $\phi < 1$, the chains of magnetized grains are divided due to the presence of glass beads.}
\label{fig:model}
\end{figure}

Let us propose a simple model to explain the scaling for both compaction parameters $\tau$ and $\eta_{\infty}$. Due to the orientational nature of the interaction, the magnetic field induces the formation of vertical chains inside the packing. The vertical length of a chain $\chi_{||} \sim Bo$ is the key physical ingredient for describing the local grains organization in the packing (see Figure \ref{fig:model}). The geometry of a chain looks like a disordered aggregate (similarily to 1d random walk) rather than a linear needle. The ``thickness" $\chi_{\perp}$ of random aggregates is given by $\chi_{\perp} \sim \sqrt{\chi_{||}} \sim \sqrt{Bo}$ as for 1d random walk. The packing of such random objects involves a loose packing fraction $\eta_0$. For increasing the packing fraction $\eta$, the structure of the disordered chains should change in order to align the grains. Indeed, the space occupied by a chain is strongly decreased.  Even if the height of the chain may increase, several chains can arrange in the space occupied by a single chain before the compaction process.  The characteristic time needed to align chains along the magnetic field, being the characteristic time $\tau$ for the packing reorganisation, is proportional to $\chi_{\perp}$, giving the right scaling $\tau \sim \sqrt{Bo}$ obtained in Figure \ref{fig:tauVsBo}. Adding the appropriate limit conditions, one writes

\begin{equation}
\tau = \tau_0 + a \sqrt{Bo}.
\label{eq:tauVSBo}
\end{equation} The experimental data are well fitted by the above square root law (see Figure \ref{fig:tauVsBo}) with the only one fitting parameter $a = 3.8 \pm 0.1$.

If one consider that the packing is an assembly of cells illustrated in Figure \ref{fig:model}, the packing fraction $\eta$ of the pile is the ratio between the volume of the grains in the cell $V_{grains}$ and the volume of the cell $V_{cell}$. One obtain

\begin{equation}
\eta = \frac{4 N \pi d^3}{3 \chi_{||}\chi_{\perp}^2},
\end{equation} where $N$ is the number of grains in the chain and $d$ is the grains diameter. Since $N \sim Bo$, $\chi_{\perp} \sim \sqrt{Bo}$ and $\chi_{||} \sim Bo$, we get the scaling $\eta \sim Bo^{-1}$. Adding the limit conditions, one writes

\begin{equation}
\eta_{\infty} = \eta_{\infty,Bo=\infty} + \frac{\eta_{\infty,Bo=0} - \eta_{\infty,Bo=\infty}}{1+Bo}.
\label{eq:rhoInfVSBo}
\end{equation} The decrease of $\eta_{\infty}$ with $Bo$ is well fitted by this relation (see Figure \label{ref:tauVsBo}). The only fitting parameter is the asymptotic packing fraction obtained for infinite Bond number $\eta_{\infty,Bo=\infty} = 0.388 \pm 0.001$.

In summary, we have derived a simple model for describing the compaction dynamics of a magnetized packing. It appears that the packing fraction as well as compaction dynamics are understood when considering the mesoscopic level. Indeed, both size and shape of particle aggregates play a major role in such system.


\section{Mixture cases}
\subsection{Observations}
The characteristic compaction time $\tau$ and the asymptotic packing fraction $\eta_{\infty}$ has been measured for mixtures of glass and metallic beads. Eight volume fractions $\phi$ of metallic beads in the mixture have been considered. Moreover, the measurements have been repeated for different values of the Bond number $Bo$. 

\begin{figure}
  \includegraphics[scale=0.55]{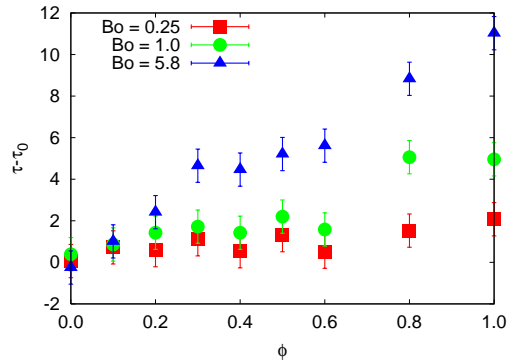}
  \caption{Evolution of the characteristic time $\tau$ as a function of the fraction of metallic beads in the mixture for different values of the Bond number $Bo$. The characteristic time $\tau_0$ obtained when $Bo = 0$ has been substracted. Each point corresponds to an average over three experiments. Error bars are indicated.}
\label{fig:tauNormVSpcFe}
\end{figure}

The figure \ref{fig:tauNormVSpcFe} presents the evolution of the characteristic time $\tau$ as a function of the fraction $\phi$ of metallic beads in the mixture for three values of the Bond number $Bo$. For each value of $\phi$, the characteristic time $\tau_0$ obtained when $Bo = 0$ has been subtracted. As expected, the compaction characteristic time $\tau$ increases with the fraction $\phi$ of metallic grains in the pile. Indeed, the cohesion inside the packing is expected to strengthen when the fraction of magnetized beads increases. This effect is amplified when the Bond number $Bo$ increases. 

\begin{figure}
  \includegraphics[scale=0.55]{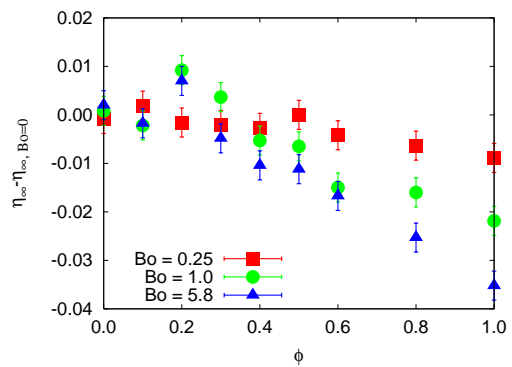}
  \caption{Evolution of the asymptotic packing fraction $\eta_{\infty}$ as a function of the fraction of metallic beads in the mixture for different values of the Bond number $Bo$. The asymptotic packing fraction $\eta_{\infty,0}$ obtained when $Bo = 0$ has been substracted. Each point corresponds to an average over three experiments. Error bars are indicated.}
\label{fig:rhoInfNormVSpcFe}
\end{figure}

The evolution of the asymptotic packing fraction $\eta_{\infty}$ as a function of the fraction $\phi$ of metallic beads in the mixture for different values of the Bond number $Bo$ is presented in Figure \ref{fig:rhoInfNormVSpcFe}. The asymptotic packing fraction $\eta_{\infty,0}$ obtained when $Bo = 0$ has been subtracted for each value of $\phi$. The packing fraction is found to decreases with the fraction $\phi$ of metallic beads. However, when $Bo>1$, a local maximum is observed for $\phi \sim 0.2$. The signification of this maximum will be discussed bellow.

\subsection{Modeling}

The previous model can be generalized to mixtures. The presence of glass beads in the packing decreases the length of the magnetic chains inside the packing (see Figure \ref{fig:model}). The decrease of the chain length as a function of the fraction $\phi$ can be evaluated simply. If we consider a reservoir of metallic and glass balls with a fraction $\phi$ of metallic balls, the probability to pick randomly a metallic ball is $\phi$. The probability to obtain $n$ successive metallic balls is $\phi^n$. In our system, when $\phi = 1$, the chains have a finite size. If $N_{max}$ is the maximum chain size, the average size $<N>$ is

\begin{equation}
<N> = \sum_{i=1}^{N_{max}} i \phi^i 
\end{equation} 

Therefore, both Equations (\ref{eq:tauVSBo}) and  (\ref{eq:rhoInfVSBo}) can be generalized in the case of magnetic and non-magnetic grains mixing. Indeed, in the case of mixtures, the characteristic size $\chi_{||}$ is proportional to $\sum_{i=1}^{N_{max}} i \phi^i Bo$. Then, we obtain

\begin{equation}
\tau = \tau_0 + a \sqrt{\frac{\sum_{i=1}^{N_{max}} i \phi^i}{\sum_{i=1}^{N_{max}} i }Bo}
\label{eq:tauVSBoMixture}
\end{equation} and

\begin{equation}
\eta_{\infty} = \eta_{\infty,Bo=\infty} + \frac{\eta_{\infty,Bo=0} - \eta_{\infty,Bo=\infty}}{1+ \frac{\sum_{i=1}^{N_{max}} i \phi^i}{\sum_{i=1}^{N_{max}} i} Bo}.
\label{eq:rhoInfVSBoMixture}
\end{equation} The division by the arithmetic series $\sum_{i=1}^{N_{max}} i$ has been used to recover the Equations (\ref{eq:tauVSBo}) and (\ref{eq:rhoInfVSBo}) when $\phi=1$. Both series could be evaluated analytically. 



\begin{figure}
  \includegraphics[scale=0.55]{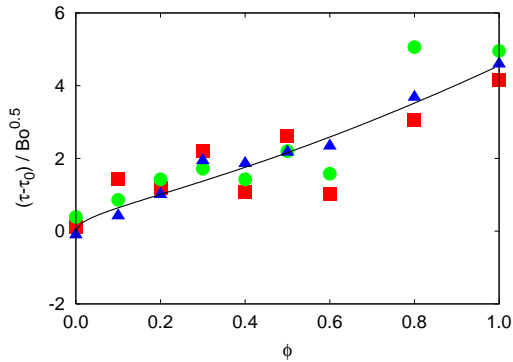}
  \caption{Scaling of the data presented in Figure \ref{fig:tauNormVSpcFe}. The plain squares, circles and triangles correspond respectively to $Bo = 0.25$, $Bo = 1$ and $Bo = 5.8$. The data collapse on a same trends and are well fitted by the Equation (\ref{eq:tauVSBoMixture}).}
\label{fig:tauScalingVSpcFe}
\end{figure}

The Figure \ref{fig:tauScalingVSpcFe} shows the scaling $(\tau-\tau_0)/\sqrt{Bo}$ as a function of $\phi$. The points corresponding to different values of the Bond numbers $Bo$ collapse on a same trends. This collapse confirms the relation $\tau \sim \sqrt{Bo}$. Moreover, the data are well fitted by the Equation (\ref{eq:tauVSBoMixture}) with $a = 4.55 \pm 0.7$ and $N_{max} = 3 \pm 1$ as free fitting parameters. 

\begin{figure}
  \includegraphics[scale=0.55]{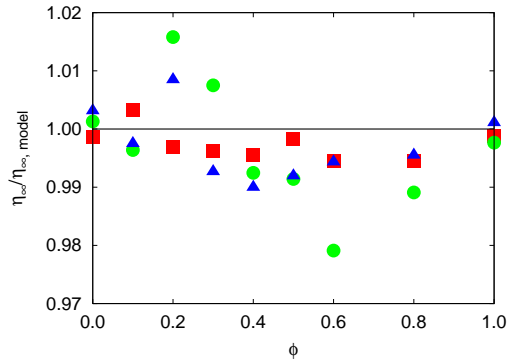}
  \caption{Deviation between the experimental values of $\eta_{\infty}$ and the prediction of the model (Eq. (\ref{eq:rhoInfVSBoMixture})). The plain squares, circles and triangles correspond respectively to $Bo = 0.25$, $Bo = 1$ and $Bo = 5.8$.}
\label{fig:rhoInfScalingVSpcFe}
\end{figure}

The deviation between the experimental values of $\eta_{\infty}$ and the prediction of the model $\eta_{\infty,model}$ obtained from the Eq. (\ref{eq:rhoInfVSBoMixture}) is presented in Figure \ref{fig:rhoInfScalingVSpcFe}. The ratio $\eta_{\infty}/\eta_{\infty,model}$ is close to unity except for the factions $\phi$ of metallic grains situated between 0.1 and 0.3 where a local maximum of $\eta_{\infty}$ is observed. Within this range, the metallic grains are mainly in contact with glass grains if the mixture is homogenous. However, during the compaction process, some clusters of metallic grains could be created in the packing. The existence of this segregation process could explain this increases of the asymptotic packing fraction. 

In summary, the simple model proposed in the pure case has been generalized to the mixture case. We expect that the magnetic chains are divided due to the presence of glass beads in the packing. 


\section{Conclusions}

The characteristic compaction time $\tau$ and the asymptotic packing fraction $\eta_{\infty}$ has been measured for mixtures of glass and metallic beads in a magnetic field. When the pile is made of metallic grains, the characteristic compaction time $\tau$ increase as $\sqrt{Bo}$ and the asymptotic packing fraction $\eta_{\infty}$ is proportional to $Bo^{-1}$. We have derived a simple mesoscopic models that explains these scalings. This model has been generalized to mixtures. 

These fundamental results should be useful to interpret the measurements made with cohesive powders used in the industry. Indeed, the tapping measurement is a very common tool to characterize the physical properties of powders.


\section{Acknowledgements}

GL and SD would like to thank FNRS for financial support. This work has been supported by INANOMAT project (IAP P6/17) of the Belgian Science Policy. The authors thanks C. Becco, F. Boschini, H. Caps, F. Ludewig, O. Gerasimov and E. Mersch for valuable discussions and J.-C. Remy for the technical support.

\end{document}